\documentclass[pra,aps,showpacs]{revtex4}
\usepackage{bm}
\usepackage{amsmath,amssymb}
\usepackage{graphicx}

\begin{document}

\title{Properties of metastable alkaline-earth-metal atoms \\
calculated using an accurate effective core potential}
\author{Robin Santra}
\author{Kevin V. Christ}
\author{Chris H. Greene}
\affiliation{Department of Physics and JILA, University of Colorado,
Boulder, CO 80309-0440}
\date{\today}
\begin{abstract}
The first three electronically excited states in the alkaline-earth-metal
atoms magnesium, calcium, and strontium comprise the 
$(nsnp)\; \phantom{}^{3}P^o_J$ ($J=0,1,2$) fine-structure manifold. All 
three states are metastable and are of interest for optical atomic clocks
as well as for cold-collision physics. An efficient technique---based on
a physically motivated potential that models the presence of the ionic
core---is employed to solve the Schr\"odinger equation for the two-electron
valence shell. In this way, radiative lifetimes, laser-induced clock shifts, 
and long-range interaction parameters are calculated for metastable
Mg, Ca, and Sr.
\end{abstract}
\maketitle

\section{Introduction}
\label{sec1}

Effectively, alkaline-earth-metal atoms are two-electron atoms,
at least at low excitation energies. Alkaline-earth-metal atoms
display---like true two-electron systems---spin singlet and triplet
level manifolds, which give rise to narrow intercombination transitions.
However, since all alkaline-earth-metal atoms possess a closed $K$ shell,
their level structure does also show notable differences to 
that of heliumlike systems. 

\begin{table}
\caption[]{The excitation energies, in eV, of the first three excited states
in the alkaline-earth-metal atoms magnesium ($3s3p$), calcium ($4s4p$), and
strontium ($5s5p$). The data are taken from Ref.~\cite{Moor71}.}
\label{tab0}
\begin{tabular}{cccc}
\hline
\hline
& Mg & Ca & Sr  \\
\hline
$^3P^o_0$ & 2.71 & 1.88 & 1.78  \\
$^3P^o_1$ & 2.71 & 1.89 & 1.80 \\
$^3P^o_2$ & 2.72 & 1.90 & 1.85 \\
\hline
\hline
\end{tabular}
\end{table}

In this work we focus on the lowest $^3P^o$ manifold, with fine-structure
components $J=0,1,2$, in the alkaline-earth-metal atoms magnesium, calcium,
and strontium. These three fine-structure states represent the first three 
electronically excited states in these species. The excitation energies, 
relative to the $^1S_0$ ground state, are reproduced in 
Table~\ref{tab0}~\cite{Moor71}.
The table illustrates that for all three atoms, the excitation energies 
correspond to optical wavelengths accessible by modern laser sources. However,
that does not imply the fine-structure components of the $^3P^o$ manifold are 
easy to excite. The transition from the ground state to the $^3P^o_1$ is 
spin-forbidden, that to the $^3P^o_0$ and the $^3P^o_2$ states is, 
additionally, in general electric-dipole-forbidden. (See Ref.~\cite{MaMo78} 
for a thorough introduction to the topic of forbidden transitions in one- and 
two-electron atoms.) This makes the $^3P^o$ manifold in the 
alkaline-earth-metal atoms interesting for
several applications in the physics of cold atoms.

The unique level structure triggered advances in laser-cooling technology
and allowed to explore processes in a magneto-optical trap at a new 
level~\cite{KaId99,BiWi01,CuOa01,GrHe02,MaTh02,XuLo02,MaCa03,XuLI03}.
Current successful approaches to an optical frequency standard using 
neutral, laser-cooled atoms are based on the relatively small linewidth
of the $^1S_0 \rightarrow \phantom{}^3P^o_1$ transition in 
calcium~\cite{OaBo99,UdDi01,WiBi02}. Significant improvement in accuracy is 
expected from the strategy of storing fermionic $^{87}$Sr atoms in an optical 
lattice and probing the $^1S_0 \rightarrow \phantom{}^3P^o_0$ 
transition~\cite{IdKa03,CoQu03,KaTa03,TaKa03,ChYe03}.

Due to the prevalence of stable alkaline-earth-metal isotopes with 
vanishing nuclear angular momentum, it is possible to investigate 
cold collisions and to perform photoassociation spectroscopy free of
any complications from hyperfine 
structure~\cite{DiVo99,ZiBi00,MaJu01,MaJu02,CaSc03}. However, it is
not possible to form a quantum-degenerate gas~\cite{CoWi02,Kett02,PeSm02} 
of ground-state alkaline-earth-metal atoms by employing evaporative cooling 
in a pure magnetic trap. Therefore, it has been suggested to make use of
the long-lived $^3P^o_2$ state~\cite{Dere01,LoBo02}. Several groups
have already succeeded in magnetically trapping $^3P^o_2$
alkaline-earth-metal atoms~\cite{KaId01,NaSi03,XuLo03,HaMo03}.
Theoretical studies have uncovered that anisotropic interactions
between $^3P^o_2$ atoms in an external magnetic field lead to 
unexpectedly rich collision physics~\cite{DePo03,SaGr03,KoSa03}.

It is the purpose of this paper to provide a concise overview of the
properties of Mg, Ca, and Sr in the $^3P^o_0$, $^3P^o_1$, and $^3P^o_2$
states. In Sec.~\ref{sec2} we describe our specific approach to the
electronic-structure problem of effective two-electron atoms. The method 
adopted is not only numerically efficient, it can 
supply atomic data at a level of accuracy comparable with high-level
{\em ab initio} calculations. Section~\ref{sec3} is devoted to a 
discussion of spontaneous decay of the individual fine-structure
components of the $^3P^o$ manifold. Then, in Sec.~\ref{sec4}, we calculate
the effect of the dynamic Stark shift on the 
$^1S_0 \rightarrow \phantom{}^3P^o_0$ clock transition. Parameters 
characterizing the long-range interaction between metastable 
alkaline-earth-metal atoms are the subject of Sec.~\ref{sec5}. We conclude
with Sec.~\ref{sec6}. Unless indicated otherwise, atomic units are used 
throughout.

\section{Treatment of the electronic-structure problem}
\label{sec2}

Our strategy toward solving the electronic-structure problem of 
alkaline-earth-metal atoms involves the following 
steps~\cite{Gree90,GrAy91,AyGr96}. We first treat the effective 
one-electron eigenvalue problem of the singly-charged alkaline-earth-metal ion
with a single valence electron. This valence electron moves in a one-particle
model potential that reproduces the valence excitation energies of the 
monocation. We represent the radial degree of freedom of the electronic
wavefunction in a finite-element basis set. From the solutions of the
one-electron problem, two-electron basis functions are constructed. The
effective two-electron Hamiltonian, which describes the neutral atom and
which fully incorporates valence-electron correlation, is represented in 
this basis and diagonalized. In this way, eigenenergies and eigenvectors 
of the two-electron valence shell are obtained.

\subsection{One-particle Hamiltonian}
\label{sec2.1}

When treating the effective one-electron problem of the monocation, we
employ the one-electron Hamiltonian
\begin{equation}
\label{s2s1e1}
h_1 = -\frac{1}{2}\nabla^2 + V + V^{(\mathrm{so})}.
\end{equation}
The operator $V$ represents the electrostatic field generated by the
noble-gaslike core. This is a central field. However, an explicit
dependence of the associated potential on the orbital angular momentum
quantum number $l$ must be included to take account of the fact that
electrons with $l=0$ dive deep into the core and experience 
relativistic effects (orbital contraction). Such scalar relativistic 
effects decrease with the average distance from the nucleus and thus 
with $l$. Therefore, we assume that the spin-angular representation of 
$V$ can be written as
\begin{equation}
\label{s2s1e2}
\left<j,m_j,l,s\right|V\left|j',m_{j'},l',s\right> = 
\delta_{j,j'}\delta_{m_j,m_{j'}}\delta_{l,l'} V_{l}(r).
\end{equation}
The quantum numbers $j$ and $m_j$ refer to the total angular momentum of
the valence electron, $s=1/2$ denotes its spin.
In this work, the following parametrization of $V_{l}(r)$ is used:
\begin{widetext}
\begin{equation}
\label{s2s1e3}
V_{l}(r) = -\frac{1}{r}\left\{2 + (Z-2)\exp{(-\alpha_{l,1}r)}
+ \alpha_{l,2} r \exp{(-\alpha_{l,3}r)}\right\} 
- \frac{\alpha_{\mathrm{cp}}}{2 r^{4}}
\left\{1 - \exp{[-(r/r_{l})^{6}]}\right\}. 
\end{equation}
\end{widetext}
$Z$ is the nuclear charge. The parameters $\alpha_{l,1}$, $\alpha_{l,2}$, 
$\alpha_{l,3}$, $\alpha_{\mathrm{cp}}$, and $r_{l}$ are listed in
Ref.~\cite{AyGr96}. The potential $V_{l}(r)$ describes the interaction of the
valence electron with the ionic core at various length scales. For very
large distances from the core, the electron is attracted by a point charge 
of charge $+2$. As the electron comes closer, the ionic core responds
to the presence of the electron and becomes polarized, as expressed by the
term proportional to $\alpha_{\mathrm{cp}}/r^4$. Below $r=r_{l}$, the electron
dives into the core. The parameters $\alpha_{l,1}$, $\alpha_{l,2}$, and 
$\alpha_{l,3}$ mediate the transition from the exterior region of the core
to the interior, where at very small length scales the electron interacts
with the unscreened charge $Z$.

Spin-orbit interaction is represented in Eq.~(\ref{s2s1e1}) by the operator 
$V^{(\mathrm{so})}$:
\begin{equation}
\label{s2s1e4}
\left<j,m_j,l,s\right|V^{(\mathrm{so})}\left|j',m_{j'},l',s\right> =
\delta_{j,j'}\delta_{m_j,m_{j'}}\delta_{l,l'} V^{(\mathrm{so})}_{l,j}(r),
\end{equation}
where
\begin{eqnarray}
\label{s2s1e5}
V_{l,j}^{(\mathrm{so})}(r) & = & \left\{j(j+1)-l(l+1)-s(s+1)\right\} \\
& & \times \frac{\alpha^2}{4 r}\frac{\mathrm{d} V_{l}}{\mathrm{d}r}
\left[1-\frac{\alpha^2}{2}V_{l}\right]^{-2}, \nonumber
\end{eqnarray}
and $\alpha \approx 1/137.036$ is the fine-structure constant.

The eigenstates of $h_1$ can now be written as 
\begin{equation}
\label{s2s1e6}
\psi = \frac{u_{l,j}(r)}{r}\left|j,m_j,l,s\right>,
\end{equation}
and we can focus on solving the radial equation
\begin{equation}
\label{s2s1e7}
\bar{h}_{l,j} u_{l,j}(r) = \varepsilon_{l,j} u_{l,j}(r), 
\enspace u_{l,j}(0) = 0,
\end{equation}
\begin{equation}
\label{s2s1e8}
\bar{h}_{l,j} = -\frac{1}{2}\frac{d^2}{dr^2} + \frac{l(l+1)}{2r^2}
+ V_{l}(r) + V_{l,j}^{(\mathrm{so})}(r).
\end{equation}
To this end, we apply a technique based on finite 
elements~\cite{Bath76,BaWi76,BrSc93,AcSh96,ReBa97,MeGr97}. 

\subsection{Finite-element basis}
\label{sec2.2}

For the radial degree of freedom, we introduce a quadratically spaced grid 
of $N+1$ grid points between $r=0$ and $r=r_{\mathrm{max}}$, i.e.,
$r_{i} = r_{\mathrm{max}} i^2/N^2$, $i=0,...,N$. In each interval 
$[r_{i},r_{i+1}]$, six unique, linearly independent fifth-order Hermite 
interpolating polynomials can be constructed that satisfy the boundary 
conditions
\begin{eqnarray}
\label{s2s2e1}
\left.\frac{d^k P_{j,i}(r)}{dr^k}\right|_{r=r_{i}} & = & \delta_{j,k},
\\
\left.\frac{d^k P_{j,i}(r)}{dr^k}\right|_{r=r_{i+1}} & = & 0,
\nonumber \\
\left.\frac{d^k Q_{j,i}(r)}{dr^k}\right|_{r=r_{i+1}} & = & \delta_{j,k},
\nonumber \\
\left.\frac{d^k Q_{j,i}(r)}{dr^k}\right|_{r=r_{i}} & = & 0,
\nonumber
\end{eqnarray}
where $j,k=0,1,2$. The functions are defined to vanish everywhere 
outside $[r_{i},r_{i+1}]$. Using the definitions
\begin{eqnarray}
\label{s2s2e2}
g_0(r) & = & f(r)^3 \left[6 f(r)^2 - 15 f(r) + 10\right], \nonumber \\
g_1(r) & = & \Delta r f(r)^3 \left[3 f(r)^2 - 7 f(r) + 4\right], \\
g_2(r) & = & \frac{1}{2} {\Delta r}^2 f(r)^3 \left[f(r) - 1\right]^2,
\nonumber
\end{eqnarray}
we have, for $r \in [r_{i},r_{i+1}]$ and $j=0,1,2$,
\begin{equation}
\label{s2s2e3}
P_{j,i}(r) = g_j(r), \enspace \Delta r = r_{i+1} - r_{i},
\enspace f(r) = -\frac{r-r_{i+1}}{\Delta r}
\end{equation}
and
\begin{equation}
\label{s2s2e4}
Q_{j,i}(r) = g_j(r), \enspace \Delta r = r_{i} - r_{i+1},
\enspace f(r) = -\frac{r-r_{i}}{\Delta r}.
\end{equation}
These formulas correct some misprints in Eq.~(49) of Rescigno 
{\em et al.}~\cite{ReBa97}.
Three continuous basis functions, with continuous first and second derivatives,
can now be associated with grid point $r_{i}$ ($i=1,...,N-1$):
\begin{equation}
\label{s2s2e5}
B_{j,i}(r) = P_{j,i}(r) + Q_{j,i-1}(r), \enspace j=0,1,2.
\end{equation}
Using these, the function value at grid point $r_{i}$, the 
first derivative, and the second derivative of any wavefunction 
can be represented. At the end points, $r_{0}$ and $r_{N}$, we set 
\begin{equation}
\label{s2s2e6}
B_{0,0}(r)=0, \enspace B_{j,0}(r) = P_{j,0}(r), \enspace j=1,2,
\end{equation}
which is consistent with the boundary condition specified in 
Eq.~(\ref{s2s1e7}), and
\begin{equation}
\label{s2s2e7}
B_{0,N}(r)=0, \enspace B_{j,N}(r) = Q_{j,N-1}(r), \enspace j=1,2,
\end{equation}
which selects solutions to Eq.~(\ref{s2s1e7}) that vanish at 
$r=r_{\mathrm{max}}$. 

The solutions to the eigenvalue problem of $\bar{h}$,
\begin{equation}
\label{s2s2e8}
\bar{h} u_{k}(r) = \varepsilon_{k} u_{k}(r),
\end{equation}
(the explicit dependence on angular momentum quantum numbers is suppressed) 
are found by expanding the eigenfunctions in terms of the 
finite-element basis functions $B_{j,i}(r)$:
\begin{equation}
\label{s2s2e9}
u_{k}(r) = \sum_{i=0}^{N} \sum_{j=0}^{2} \beta_{(j,i),k}
B_{j,i}(r).
\end{equation}
Hence, we have to solve the generalized eigenvalue problem
\begin{equation}
\label{s2s2e10}
\bar{\bm h} {\bm \beta}_k = \varepsilon_k 
{\bm o} {\bm \beta}_k,
\end{equation}
where 
\begin{equation}
\label{s2s2e11}
\left(\bar{\bm h}\right)_{(j,i),(j',i')} = 
\int_{0}^{r_{\mathrm{max}}} dr B_{j,i}(r) \bar{h} B_{j',i'}(r),
\end{equation}
\begin{equation}
\label{s2s2e12}
\left({\bm o}\right)_{(j,i),(j',i')} =
\int_{0}^{r_{\mathrm{max}}} dr B_{j,i}(r) B_{j',i'}(r),
\end{equation}
and 
\begin{equation}
\label{s2s2e13}
\left({\bm \beta}_k\right)_{(j,i)} = \beta_{(j,i),k}.
\end{equation}
Since the basis function $B_{j,i}(r)$ vanishes outside $[r_{i-1},r_{i+1}]$,
both matrices, $\bar{\bm h}$ and ${\bm o}$, have a simple banded structure 
with small bandwidth. The high degree of sparsity is ideal for iterative
solvers~\cite{GoLo96}. Using the Lanczos-based package ARPACK~\cite{LeSo98},
we calculate---for each valid combination of angular momentum quantum numbers
$l$ and $j$---the first 18 eigenfunctions outside the core shells. 
In other words, solutions associated with inner shells are skipped. Thus,
the selected valence-electron solutions display correct nodal behavior.

\subsection{Two-particle Hamiltonian}
\label{sec2.3}

We describe the neutral atom employing the effective two-electron Hamiltonian
\begin{equation}
\label{s2s3e1}
h_2 = h_1(1) + h_1(2) + \frac{1}{|{\bm x}_1 - {\bm x}_2|};
\end{equation}
i.e., within the approach we take, both valence electrons move in the
field of the model potential $V$ (Eqs.~(\ref{s2s1e2}) and (\ref{s2s1e3})),
are subject to spin-orbit coupling through $V^{(\mathrm{so})}$ 
(Eqs.~(\ref{s2s1e4}) and (\ref{s2s1e5})), and experience mutual Coulomb 
repulsion.

In order to tackle the eigenvalue problem of $h_2$, we construct two-electron 
basis functions, in $jj$ coupling, from the solutions of the one-electron 
problem:
\begin{widetext}
\begin{equation}
\label{s2s3e2}
\Phi = {\cal A} \frac{u_{l_1,j_1}(r_1)}{r_1} \frac{u_{l_2,j_2}(r_2)}{r_2}
\sum_{m_{j_1},m_{j_2}} C(j_1 j_2 J; m_{j_1} m_{j_2} M)
\left|j_1,m_{j_1},l_1,s_1\right>\left|j_2,m_{j_2},l_2,s_2\right>, 
\end{equation}
\end{widetext}
where ${\cal A}$ is an antisymmetrization operator and 
$C(j_1 j_2 J; m_{j_1} m_{j_2} M)$ is a Clebsch-Gordan coefficient.
The total electronic angular momentum, $J$, the associated projection, 
$M$, as well as the atomic parity, $\Pi=(-1)^{l_1+l_2}$, are the only 
good quantum numbers in the two-electron problem, Eq.~(\ref{s2s3e1}). The 
matrix representation of $h_1(1) + h_1(2)$---the Hamiltonian of two 
noninteracting valence electrons---in the basis of the two-electron 
functions is, of course, diagonal. Using the familiar relation~\cite{Jack98} 
\begin{equation}
\label{s2s3e3}
\frac{1}{|{\bm x}_1 - {\bm x}_2|} =  
\sum_{l=0}^{\infty} \sum_{m=-l}^{l} \frac{4\pi}{2l+1} 
\frac{r_{<}^l}{r_{>}^{l+1}} Y_{lm}(\vartheta_1,\varphi_1) 
Y_{lm}^{\ast}(\vartheta_2,\varphi_2),  
\end{equation}
\[
r_{<} = {\mathrm{min}}\{r_1,r_2\}, \enspace 
r_{>} = {\mathrm{max}}\{r_1,r_2\},
\]
together with standard Wigner--Racah algebra~\cite{Edmo57,FaRa59,Zare88}, 
the matrix of the electron--electron interaction term in Eq.~(\ref{s2s3e1})
can be calculated.

The resulting matrix representation of $h_2$ is sufficiently small to 
allow for full numerical diagonalization. Properties of the valence shell 
can now be computed from the spectrum and eigenvectors of this matrix.
In order to estimate the accuracy of our calculations, we first 
calculate the physical quantity of interest (a radiative decay rate, for 
example), ensuring convergence with respect to all basis set parameters. In 
this step, the eigenenergies and the relevant matrix elements with respect 
to the eigenvectors are taken from our calculations. Then we recalculate 
that physical quantity by replacing the {\em energies} of the most 
important states by experimental energies taken from Ref.~\cite{Moor71}. 
Finally, we quote the result based on the experimental energies together 
with an error estimate given by the difference from the result calculated 
using theoretical energies. 

\section{Radiative decay rates}
\label{sec3}

\subsection{$^3P^o_0$}
\label{sec3.1}

The bosonic atoms $^{24}$Mg, $^{40}$Ca, and $^{88}$Sr are by far the most 
abundant isotopes of magnesium, calcium, and strontium, respectively. The 
nuclei in these isotopes have vanishing nuclear spin, implying that the 
quantum number $J=0$ in the $^3P^o_0$ state is exact. As all one-photon 
operators have a rank greater than zero, the most important way the 
$^3P^o_0$ state can decay is via an E1M1 two-photon process, which, in 
contrast to E1E1, can mediate the parity change between $^3P^o_0$ and 
$^1S_0$. The E1M1 process has been analyzed in detail for the radiative 
decay of the $^3P^o_0$ state in heavy heliumlike ions~\cite{Drak85,SaJo02}, 
where this process competes with the usual E1 decay. (In heliumlike systems, 
the $^3P^o_0$ state is not the first excited state.) 

The E1M1-decay rate from the $^3P^o_0$ state, symbolized by the
state vector $\left|i\right>$ with energy $E_i$, to the $^1S_0$ ground state
($\left|f\right>$, $E_f$) can be readily derived within the framework of 
quantum electrodynamics~\cite{CrTh98} and is given by 
\begin{eqnarray}
\label{s3s1e1}
\Gamma_{\mathrm{E1M1}} & = & \frac{8}{27 \pi} \alpha^6 
\int_{0}^{\infty} d\omega_1 \omega_1^3 
\int_{0}^{\infty} d\omega_2 \omega_2^3 \nonumber \\
& \times & 
\left| \sum_{n_{+}}
\frac{\left<f\right.\parallel M_1 \parallel \left.n_{+}\right>
\left<n_{+}\right. \parallel D \parallel \left.i\right>}{E_{n_{+}}+
\omega_1-E_i}
\right. \nonumber \\
& & + \left. \sum_{n_{-}}
\frac{\left<f\right. \parallel D \parallel \left.n_{-}\right>
\left<n_{-}\right.\parallel M_1 \parallel \left.i\right>}{E_{n_{-}}+
\omega_2-E_i}
\right|^2 \nonumber \\
& \times & \delta(E_f+\omega_1+\omega_2-E_i).
\end{eqnarray}
We use the notation 
$\left<\cdot\right.\parallel\cdot\parallel\left.\cdot\right>$ to denote 
a {\em reduced} matrix element~\cite{Edmo57,FaRa59,Zare88}. 
Equation~(\ref{s3s1e1}) illustrates that the two emitted photons with energies
$\omega_1$ and $\omega_2$, respectively, share the available energy,
$E_i-E_f$. The decay can proceed in two indistinguishable ways. 
In one of the two, the E1 photon is emitted first, mediated by the 
electric-dipole operator 
\begin{equation}
\label{s3s1e2}
D_q = - \sum_{i} r_{i} C_{1,q}(\vartheta_{i},\varphi_{i}), 
\end{equation} 
and a virtual transition to the manifold of states $\left|n_{+}\right>$ 
with even parity and $J=1$ takes place. (We use the definition 
$C_{l,q} = \sqrt{4 \pi/2 l + 1} Y_{l,q}$.)
The intermediate states are then coupled to the ground state
by the magnetic-dipole operator 
\begin{equation}
\label{s3s1e3}
M_{1,q} = -\mu_B \left(J_q+S_q\right), 
\end{equation}
$\mu_B$ denoting the Bohr magneton. Alternatively, we first
have a virtual M1 transition to states $\left|n_{-}\right>$ 
with odd parity and $J=1$,
followed by E1 emission. In the case of the lowest $^3P^o_0$ state in Mg,
Ca, and Sr, the energy denominators in Eq.~(\ref{s3s1e1}) differ from zero 
for all $\omega_1$ and $\omega_2$, so that there are no numerical 
complications arising from poles. 

\begin{table}
\caption[]{Spontaneous decay rate, in s$^{-1}$, of the $^3P^o_0$ state in the
alkaline-earth-metal atoms magnesium ($3s3p$), calcium ($4s4p$), and
strontium ($5s5p$). The spinless bosonic isotopes can decay only via E1M1
two-photon emission. The stable fermionic isotopes---$^{25}$Mg, $^{43}$Ca, and
$^{87}$Sr---decay by a hyperfine-induced E1 one-photon process.}
\label{tab1}
\begin{tabular}{cccc}
\hline
\hline
 & Mg & Ca & Sr \\
\hline
boson & $1.6(1) \times 10^{-13}$ & $3.9(2) \times 10^{-13}$
& $5.5(3)\times 10^{-12}$  \\
fermion & $9(3) \times 10^{-4}$ & $3(1)\times 10^{-3}$ &
$9(3) \times 10^{-3}$ \\
\hline
\hline
\end{tabular}
\end{table}

Our calculated E1M1 rates are presented in Table~\ref{tab1}. They are,
not surprisingly, very small throughout. Even in strontium, which decays
faster by more than an order of magnitude than calcium and magnesium,
the lifetime is about $5800$ years. This lifetime is comparable to the 
half-life of the radioactive isotope $^{14}$C. 

The nuclear angular momentum, $I$, in the fermionic isotopes of magnesium,
calcium, and strontium differs from zero~\cite{NNDC03}: $I$($^{25}$Mg) = $5/2$,
$I$($^{43}$Ca) = $7/2$, $I$($^{87}$Sr) = $9/2$. The finite nuclear 
magnetic-dipole moment couples the $^3P^o_0$ state to states of the same 
parity but with $J=1$. The most important among these states are the 
energetically lowest $^3P^o_1$ and $^1P_1$ states. These can decay directly 
to the ground state by emission of an E1 photon.

The interaction of the valence electrons with the magnetic-dipole 
moment, ${\bm M}_1^{\mathrm{nuc}}$, and the electric-quadrupole moment,
${\bm Q}_2^{\mathrm{nuc}}$, of the atomic nucleus is given by the 
hyperfine-coupling operator~\cite{Sobe92}
\begin{widetext}
\begin{eqnarray}
\label{s3s1e4}
V_{\mathrm{hf}} & = & -\sqrt{3} \alpha \sum_i
\left\{\frac{8 \pi}{3} \delta({\bm x}_i) 
[{\bm s}_i\otimes{\bm M}_1^{\mathrm{nuc}}]_{0,0} 
+ \frac{1}{r_i^3} [{\bm l}_i \otimes {\bm M}_1^{\mathrm{nuc}}]_{0,0} 
-\frac{\sqrt{10}}{r_i^3} 
[[{\bm C}_2(\vartheta_i,\varphi_i)\otimes{\bm s}_i]_1\otimes
{\bm M}_1^{\mathrm{nuc}}]_{0,0}\right\} \nonumber \\
& & - \sum_i
\frac{\sqrt{5}}{r_i^3}[{\bm C}_2(\vartheta_i,\varphi_i)
\otimes{\bm Q}_2^{\mathrm{nuc}}]_{0,0}.
\end{eqnarray}
\end{widetext}
We employ the notation 
\begin{equation}
\label{s3s1e5}
[{\bm A}_{k_1}\otimes{\bm B}_{k_2}]_{K,M} 
= \sum_{m_1,m_2} C(k_1 k_2 K; m_1 m_2 M) A_{k_1,m_1} B_{k_2,m_2}
\end{equation}
for the rank-$K$ tensor product of tensors ${\bm A}_{k_1}$ (rank
$k_1$) and ${\bm B}_{k_2}$ (rank $k_2$). 

It is not {\em a priori} clear that the valence-electron wavefunctions derived 
from the model potential, Eq.~(\ref{s2s1e3}), have the correct behavior near 
the nucleus, even if valence excitation spectra can be well reproduced. In 
order to check this, we employed the hyperfine-coupling operator 
$V_{\mathrm{hf}}$ to quantitatively calculate hyperfine parameters for the 
($5s5p$) $^3P^o_1$ and $^3P^o_2$ states in $^{87}$Sr~\cite{NNDC03}. 
For $^3P^o_1$ we find A = -278 MHz and B = -30.8 MHz. This result is in 
reasonable agreement with an experiment by zu Putlitz~\cite{zuPu63}, who 
measured A = -260 MHz and B = -35.7 MHz. Heider and Brink~\cite{HeBr77} 
determined the hyperfine parameters for the $^3P^o_2$ state: 
A = -213 MHz and B = 67.2 MHz. Our calculation gives
A = -231 MHz and B = 56.5 MHz. Even though the agreement is not perfect,
it indicates that we should be able to calculate the hyperfine-induced
coupling of the $^3P^o_0$ state to other electronic levels with an 
accuracy of about $10$\%.

Since the hyperfine-coupling operator $V_{\mathrm{hf}}$ (Eq.~(\ref{s3s1e4})) is
a tensor of rank $0$ with respect to rotations of nucleus and electrons, 
it conserves the total angular momentum of the atom, $F$. Furthermore,
for the transition from $^3P^o_0$ to $^1S_0$, F=I. First-order 
perturbation theory leads to the following formula for the hyperfine-induced 
spontaneous decay rate:
\begin{eqnarray}
\label{s3s1e6}
\Gamma_{\mathrm{hf}} & = & \frac{4}{9} \frac{\alpha^3 \omega^3}{2 I + 1} \\
& & \times 
\left| \sum_{n_{-}}
\frac{\left<f\right.\parallel D \parallel \left.n_{-}\right>
\left<F,n_{-}\right.\parallel V_{\mathrm{hf}} \parallel \left.F,i\right>}{
E_{n_{-}}-E_i}
\right|^2. \nonumber 
\end{eqnarray}
Here, $\omega=E_i-E_f$ is the energy of the emitted photon, and as before,
the states $\left|n_{-}\right>$ have odd parity and $J=1$. The 
electric-quadrupole term in $V_{\mathrm{hf}}$ does not play a role, for it 
couples $^3P^o_0$ to states with $J=2$. These, however, cannot decay to the 
ground state via E1 one-photon emission. We use the abbreviated notation 
$\left|F,n\right>$ in the reduced matrix elements of $V_{\mathrm{hf}}$
in Eq.~(\ref{s3s1e6}) to indicate that these states are angular-momentum 
coupled states of the entire atom with total angular momentum $F$.

The $^3P^o_0$ decay rates in the fermionic alkaline-earth-metal isotopes,
as calculated on the basis of Eq.~(\ref{s3s1e6}), are collected in 
Table~\ref{tab1}. We note that the lifetimes are now much shorter---by 
about $10$ orders of magnitude---than in the bosonic isotopes. Nevertheless, 
this is still sufficiently long to make the 
$^1S_0$ $\rightarrow$ $^3P^o_0$ transition interesting for optical atomic 
clocks. Especially the strontium isotope $^{87}$Sr---which has a rather 
narrow $^3P^o_0$ linewidth of about $1$ mHz---is a candidate for a future 
ultraprecise atomic clock~\cite{KaTa03}. In fact, this is the only one of 
the three isotopes $^{25}$Mg, $^{43}$Ca, and $^{87}$Sr, for which we are 
aware of an analogous $^3P^o_0$ decay rate determination: 
{Pal'chikov}~\cite{Palc02} obtained a rate of $5.5 \times 10^{-3}$ s$^{-1}$
for this $^{87}$Sr transition. This is somewhat smaller than our result, 
apparently for two reasons. First, we have used experimental excitation 
energies, and second, we have included a larger number of intermediate states 
when evaluating Eq.~(\ref{s3s1e6}). If we apply the same restrictions as 
Pal'chikov, we find $6 \times 10^{-3}$ s$^{-1}$.

\subsection{$^3P^o_1$}
\label{sec3.2}

The $^3P^o_1$ state can decay via two obvious decay modes. On one hand,
M1 one-photon decay from the $^3P^o_1$ to the $^3P^o_0$ state can take place.
This mode, however, is slower by many orders of magnitude than the E1 decay 
directly to the $^1S_0$ ground state. The only reason the $^3P^o_1$ state
is metastable is the necessity of a spin flip. The E1 decay is enabled by the 
presence of spin-orbit coupling---a relativistic effect that becomes 
more pronounced in heavier atoms.

\begin{table}
\caption[]{Spontaneous decay rate, in s$^{-1}$, of the $^3P^o_1$ state in the
alkaline-earth-metal atoms magnesium ($3s3p$), calcium ($4s4p$), and
strontium ($5s5p$). The decay is dominated by the E1 rate for the transition
$^3P^o_1$ $\rightarrow$ $^1S_0$.}
\label{tab2}
\begin{tabular}{ccc}
\hline
\hline
Mg & Ca & Sr \\
\hline
$3.6(1) \times 10^{2}$ & $2.1(2) \times 10^{3}$ & $4.1(4)\times 10^{4}$  \\
\hline
\hline
\end{tabular}
\end{table}

The E1 decay rate is given 
by~\cite{Sobe92}
\begin{equation}
\label{s3s2e1}
\Gamma_{\mathrm{E1}} = \frac{4}{3} \frac{\alpha^3 \omega^3}{2 J_i + 1}
\left|\left<f\right.\parallel D\parallel\left.i\right>\right|^2.
\end{equation}
Our calculated reduced electric-dipole matrix elements for the 
$^3P^o_1 \rightarrow \phantom{}^1S_0$ transition agree to about the $10$\%
level with recent high-level many-body and configuration-interaction 
calculations~\cite{PoKo01,SaII02}. The result for the calculated decay rate,
$\Gamma_{\mathrm{E1}}$, is presented in Table~\ref{tab2}. As expected,
the decay rate increases from Mg to Sr due to enhanced spin-orbit coupling.

The $^3P^o_1$ decay rate in all three alkaline-earth-metal atoms is 
sufficiently large to allow their direct experimental determination.
The most recent measurements give $1.9(3) \times 10^{2}$ s$^{-1}$ 
for Mg~\cite{GoNo92}, $2.9(2) \times 10^{3}$ s$^{-1}$ for Ca~\cite{DrIg97},
and $4.7(1) \times 10^{4}$ s$^{-1}$ for Sr~\cite{DrIg97}.
Perfect agreement is not expected for our simple model approach.
Nonetheless, we may conclude that our calculated reduced electric-dipole 
matrix elements are apparently accurate to about the $10$\% level.

\subsection{$^3P^o_2$}
\label{sec3.3}

Alkaline-earth-metal atoms in the $^3P^o_0$ state do not possess
a magnetic-dipole moment. In the $^3P^o_1$ state they do, but
even in Mg this state decays after about $5$ ms. The third candidate,
$^3P^o_2$, is the only one useful for magnetic trapping~\cite{BeEr87},
due to its inherently long lifetime~\cite{Dere01}.

There are several different modes through which the $^3P^o_2$ state 
can decay. Transitions within the $^3P^o$ fine-structure manifold are 
mediated by parity-conserving one-photon operators. The rate associated with 
a magnetic-dipole transition to the $^3P^o_1$ state can be written 
as~\cite{Sobe92}
\begin{equation}
\label{s3s3e1}
\Gamma_{\mathrm{M1}} = \frac{4}{3} \frac{\alpha^3 \omega^3}{2 J_i + 1}
\left|\left<f\right.\parallel M_1\parallel\left.i\right>\right|^2.
\end{equation}
While it is permissible to neglect magnetic-octupole decay, the 
electric-quadrupole decay to the $^3P^o_1$ state must be taken into 
consideration. The rate for the latter process is given by~\cite{Sobe92}
\begin{equation}
\label{s3s3e2}
\Gamma_{\mathrm{E2}} = \frac{1}{15} \frac{\alpha^5 \omega^5}{2 J_i + 1}
\left|\left<f\right.\parallel Q\parallel\left.i\right>\right|^2,
\end{equation}
where
\begin{equation}
\label{s3s3e3}
Q_q = - \sum_{i} r_{i}^2 C_{2,q}(\vartheta_{i},\varphi_{i}).
\end{equation}
The E2 mechanism also allows for decay from $^3P^o_2$ to $^3P^o_0$.
Finally, direct decay to the $^1S_0$ ground state requires a 
parity-changing one-photon operator of rank $2$, i.e., the
magnetic-quadrupole operator~\cite{Mizu64,Gars67}
\begin{widetext}
\begin{equation}
\label{s3s3e4}
M_{2,q} = - \sqrt{6} \alpha \sum_{i} r_{i} 
\left\{[{\bm C}_1(\vartheta_{i},\varphi_{i})\otimes{\bm s}_i]_{2,q}
+\frac{1}{3}[{\bm C}_1(\vartheta_{i},\varphi_{i})\otimes{\bm l}_i]_{2,q}
\right\}.
\end{equation}
\end{widetext}
The associated rate is~\cite{Sobe92}
\begin{equation}
\label{s3s3e5}
\Gamma_{\mathrm{M2}} = \frac{1}{15} \frac{\alpha^5 \omega^5}{2 J_i + 1}
\left|\left<f\right.\parallel M_2\parallel\left.i\right>\right|^2.
\end{equation}

\begin{table}
\caption[]{Spontaneous decay rates, in s$^{-1}$, of the $^3P^o_2$ state in the
alkaline-earth-metal atoms magnesium ($3s3p$), calcium ($4s4p$), and
strontium ($5s5p$).}
\label{tab3}
\begin{tabular}{ccccc}
\hline
\hline
transition & type & Mg & Ca & Sr  \\
\hline
$^3P^o_2 \rightarrow \phantom{}^3P^o_1$ & M1 & $1(1) \times 10^{-6}$
& $1.6(3) \times 10^{-5}$ & $8(2) \times 10^{-4}$ \\
$^3P^o_2 \rightarrow \phantom{}^3P^o_1$ & E2 & $1(3) \times 10^{-12}$
& $3(1) \times 10^{-10}$ & $3(1) \times 10^{-7}$ \\
$^3P^o_2 \rightarrow \phantom{}^3P^o_0$ & E2 & $0(1) \times 10^{-11}$
& $9(3) \times 10^{-10}$ & $9(4) \times 10^{-7}$ \\
$^3P^o_2 \rightarrow \phantom{}^1S_0$ & M2 & $4.2(2) \times 10^{-4}$
& $1.1(1) \times 10^{-4}$ & $1.1(2) \times 10^{-4}$ \\
\hline
total rate      & & $4.2(2) \times 10^{-4}$
 & $1.3(1) \times 10^{-4}$
& $9(2) \times 10^{-4}$ \\
Ref.~\cite{Dere01} & & $4.42 \times 10^{-4}$
 & $1.41 \times 10^{-4}$
& $9.55 \times 10^{-4}$ \\
\hline
\hline
\end{tabular}
\end{table}

Our calculated decay rates are shown in Table~\ref{tab3}. We see that
in Mg and Ca, the exotic M2 mode dominates by far. In Sr, however,
due to the $\omega^3$ dependence in Eq.~(\ref{s3s3e1}), M1 decay makes
the biggest contribution. Also shown in Table~\ref{tab3} is a comparison
of our calculated total rates with those determined by Derevianko, 
Ref.~\cite{Dere01}. We find good agreement and conclude that the 
$^3P^o_2$ lifetimes in Mg, Ca, and Sr are, respectively, $2.4(1)\times 10^3$ s,
$7.7(6)\times 10^3$ s, and $1.1(3)\times 10^3$ s. The first
measurement, by Yasuda and Katori, of the $^3P^o_2$ lifetime in $^{88}$Sr
yielded $500^{+280}_{-130}$ s~\cite{YaKa03}, which is compatible with 
the current level of theory.

\section{Dynamic Stark effect}
\label{sec4}

As mentioned in the introduction, Sec.~\ref{sec1}, the transition from the
ground state to the $^3P^o_0$ state in the fermionic isotope $^{87}$Sr 
is of interest for the development of an ultraprecise optical clock 
based on neutral atoms trapped in an optical 
lattice~\cite{IdKa03,CoQu03,KaTa03,TaKa03,ChYe03}. It has been pointed out
by Katori {\em et al.} that with this approach a relative accuracy of better
than $10^{-17}$ should be achievable~\cite{KaTa03}. 

The optical lattice serves three purposes. If each lattice site is occupied
by at most one atom, collisional clock shifts can be suppressed
(see Ref.~\cite{ChYe03} for a discussion of frequency shifts caused by
dipole--dipole interactions among atoms in different lattice sites).
Furthermore, using trapped atoms, long interaction times with the probe
laser can be realized. This is particularly important when exciting 
a transition with exceptionally small linewidth (see Table~\ref{tab1}).
A third benefit is the possibility of confining the atoms in the 
Lamb--Dicke regime~\cite{Dick53,IdKa03,KaTa03,TaKa03}. In this way, atomic
recoil- and Doppler-shifts can be eliminated. 

The intense laser field needed to confine the atoms perturbs them. To lowest
order in perturbation theory, each atomic level is shifted by a level-specific
amount proportional to the intensity of the laser~\cite{Mitt93}. This effect
is referred to as {\em dynamic} (or {\em ac}) Stark shift. The dynamic
Stark shift of state $\left|m\right>$ can be calculated from~\cite{Mitt93}
\begin{eqnarray}
\label{s4e1}
\Delta E_m & = & \frac{2\pi}{3} \frac{\alpha {\cal I}}{2 J_m + 1} \\
& & \times
\sum_{n} \left\{
\frac{\left|\left<n\right.\parallel D \parallel \left.m\right>\right|^2}
{E_m-E_n-\omega}
+
\frac{\left|\left<n\right.\parallel D \parallel \left.m\right>\right|^2}
{E_m-E_n+\omega}
\right\}. \nonumber
\end{eqnarray}
In this equation, ${\cal I}$ is the laser intensity in units of
${\cal I}_0 = e^2/a_0^4 = 6.4364 \times 10^{15}$ W/cm$^2$; $\omega$ is the 
laser photon energy.

However, it is possible to find a {\em magic} wavelength for the trapping
laser field such that the dynamic Stark shifts of the two levels involved
in the clock transition are identical~\cite{IdKa03,KaTa03,TaKa03}.
In other words, at the magic wavelength, the clock frequency is virtually
unaffected by the presence of the optical lattice. In a recent calculation,
the magic wavelength for the $^1S_0 \rightarrow \phantom{}^3P^o_0$
transition in Sr was found to be about $800$~nm~\cite{KaTa03}, in 
agreement with the experimental value of $813.5(9)$~nm~\cite{TaKa03}.

\begin{table}
\caption[]{Magic wavelength, in nm, for the transition
$(ns^2)\; \phantom{}^1S_0 \rightarrow (nsnp)\; \phantom{}^3P^o_0$.
At this wavelength, the dynamic Stark shifts, Eq.~(\ref{s4e1}),
of both levels, $^1S_0$ and $^3P^o_0$, are identical.}
\label{tab4}
\begin{tabular}{ccc}
\hline
\hline
Mg & Ca & Sr \\
\hline
$470(10)$ & $700(50)$ & $770(60)$ \\
\hline
\hline
\end{tabular}
\end{table}

Using Eq.~(\ref{s4e1}), we determined the magic wavelength for the proposed
clock transition within our model-potential approach. The results are 
collected in Table~\ref{tab4}. Within our error bars, our result for Sr
is in agreement with the previous calculation and with the experiment. To our
knowledge, Mg and Ca have not been considered before. As can be seen
in Table~\ref{tab4}, the magic wavelength in Mg is too blue to be ideal for 
trapping with current laser sources. The magic wavelength we find in Ca 
appears to be much more attractive. In view of this result and of the
small decay width of the $^3P^o_0$ state in $^{43}$Ca (see Table~\ref{tab1}),
fermionic calcium appears to be an excellent candidate for an optical atomic 
clock and may serve as an interesting alternative to $^{87}$Sr.

\section{Long-range parameters}
\label{sec5}

\subsection{$^1S^0$}
\label{sec5.1}

While collisions among atoms are to be minimized in atomic clocks, interatomic
interactions at temperatures close to absolute zero are highly intriguing
and are responsible for some of the most interesting effects in the
physics of ultracold gases~\cite{PeSm02,WeBa99,DaGi99,Legg01}. 
In order to understand cold collisions, detailed knowledge about the 
long-range behavior of the interatomic interaction potential is 
indispensable~\cite{WeBa99}.

Here, we concentrate on interactions between two identical alkaline-earth-metal
atoms in the same quantum state. If the two interacting atoms have vanishing 
angular momentum, $J=0$, then at the leading, second-order level of the 
long-range expansion of the interaction energy~\cite{Marg39,DaDa66,Chan67}, 
both atoms can only become polarized---and thus reduce the overall energy of 
the system---by making virtual electric-dipole transitions to states with 
$J=1$. In this case, the electric dipole--dipole dispersion interaction 
energy between atoms in atomic state $\left|m\right>$, separated by distance 
$R$, is well known to be given by the leading $-C_6/R^6$ term, 
where~\cite{Marg39,DaDa66,Chan67}
\begin{equation}
\label{s5e1}
C_6 = \frac{2}{3} 
\sum_{n_1,n_2} 
\frac{\left|\left<n_1\right.\parallel D \parallel \left.m\right>\right|^2
\left|\left<n_2\right.\parallel D \parallel \left.m\right>\right|^2}
{E_{n_1}+E_{n_2}-2 E_m}.
\end{equation}

\begin{table}
\caption[]{$C_6$ long-range dispersion coefficient, in atomic units, for
two $(ns^2)\; \phantom{}^1S_0$ alkaline-earth-metal atoms.}
\label{tab5}
\begin{tabular}{cccc}
\hline
\hline
& Mg & Ca & Sr \\
\hline
This work & $620(5)$ & $2150(60)$ & $3260(100)$ \\
Ref.~\cite{PoDe02} & $627(12)$ & $2221(15)$ & $3170(196)$ \\
Ref.~\cite{Stan94} & $648$ & $2042$ & $3212$ \\
\hline
\hline
\end{tabular}
\end{table}

For $^1S_0$ atoms, the states $\left|n_1\right>$, $\left|n_2\right>$ have 
angular momentum $J=1$ and odd parity. The $C_6$ coefficients we calculated
for Mg, Ca, and Sr in the atomic ground state are listed in Table~\ref{tab5}.
Our numbers can be compared with high-level {\em ab initio} data from 
Refs.~\cite{PoDe02} and \cite{Stan94}. The high quality of our results is
evident. The monotonic increase of the $C_6$ coefficient from Mg to Sr
is a simple consequence of the greater polarizability of the heavier 
alkaline-earth-metal atoms.

\subsection{$^3P^o_0$}
\label{sec5.2}

\begin{table}
\caption[]{$C_6$ long-range dispersion coefficient, in atomic units, for
two $(nsnp)\; \phantom{}^3P^o_0$ alkaline-earth-metal atoms.}
\label{tab6}
\begin{tabular}{ccc}
\hline
\hline
Mg & Ca & Sr \\
\hline
$980(30)$ & $3020(200)$ & $5260(500)$ \\
\hline
\hline
\end{tabular}
\end{table}

More interesting for future experiments with metastable atoms is the
long-range physics of the $^3P^o_0$ state. As we have seen in Table~\ref{tab1},
in this state the lifetime, under spontaneous emission, of bosonic 
alkaline-earth-metal atoms is virtually unlimited. Any quenching of the
metastable state will be induced by collisions with either atoms or 
photons. It is not clear, for example, on what timescale a gas consisting
of $^3P^o_0$ atoms will disintegrate due to inelastic processes, how these 
processes can be controlled, or what role elastic collisions have to play. An 
important first step toward an answer consists again in determining the
corresponding $C_6$ coefficient. For $^3P^o_0$ atoms, the states 
$\left|n_1\right>$, $\left|n_2\right>$ in Eq.~(\ref{s5e1}) have angular 
momentum $J=1$ and even parity. The result of our calculation is
shown in Table~\ref{tab6}. As for ground state atoms, we observe a 
monotonic dependence of the $C_6$ coefficient on the atomic number.
Additionally, it should be noted that in the $^3P^o_0$ state the
$C_6$ coefficients for all three atomic species are greater than in
the ground state by a factor of about $1.5$. This expresses the fact
that for excited atoms, the relevant energy denominators in Eq.~(\ref{s5e1})
are smaller on average.

\subsection{$^3P^o_2$}
\label{sec5.3}

As Derevianko pointed out~\cite{Dere01}, the very long lifetime (see 
Sec.~\ref{sec3.3}) of alkaline-earth-metal atoms in the lowest 
$^3P^o_2$ state makes magnetic trapping of these metastable species 
experimentally practical. This expectation has turned out to be 
correct~\cite{KaId01,NaSi03,XuLo03,HaMo03}. However, what ultimately 
limit the efficiency of evaporative cooling in a magnetic trap are
inelastic collisions~\cite{KoSa03}, not radiative decay processes. The 
strong coupling to inelastic scattering channels is caused by the anisotropic 
long-range interactions between the nonspherical $^3P^o_2$ atoms.

\begin{table}
\caption[]{Atomic electric-quadrupole moment, in atomic units,
in the $(nsnp)\; \phantom{}^3P^o_2$ state.}
\label{tab7}
\begin{tabular}{cccc}
\hline
\hline
& Mg & Ca & Sr \\
\hline
This work & $8.38$ & $12.7$ & $15.4$ \\
Ref.~\cite{DePo03} & $8.46(8)$ & $12.9(4)$ & $15.6(5)$ \\
Ref.~\cite{Dere01} & $8.59$    & $13.6$    & $16.4$ \\
\hline
\hline
\end{tabular}
\end{table}

The most important parameter in this context is the atomic 
electric-quadrupole moment~\cite{Dere01,SaGr03}. For an atom in state 
$\left|m\right>$ with angular momentum $J_m$, the electric-quadrupole 
moment can be written as~\cite{SaGr03}
\begin{equation}
\label{s5e2}
{\cal Q} = 2 \sqrt{\frac{J_m (2 J_m - 1)}{(J_m + 1)(2 J_m + 1)(2 J_m + 3)}}
\left<m\right. \parallel Q \parallel \left.m\right>.
\end{equation}
${\cal Q}$ vanishes, of course, if $J_m=0$ or $J_m=1/2$. The atomic 
electric-quadrupole moment gives rise to direct, first-order 
quadrupole--quadrupole interactions proportional to $1/R^5$~\cite{Knip38}. 
Derevianko was the first to calculate ${\cal Q}$ for $^3P^o_2$ 
alkaline-earth-metal atoms~\cite{Dere01}; improved data were published 
recently~\cite{DePo03}. We list the results from Refs.~\cite{Dere01} and 
\cite{DePo03} together with our calculated values in Table~\ref{tab7}. Again, 
our model calculations provide data of quality similar to the data derived 
from high-level many-body theory, Ref.~\cite{DePo03}. The monotonic increase 
of ${\cal Q}$ with respect to atomic number reflects the increase in
spatial extension of the excited $np$ orbital, where $n=3$ for Mg,
$n=4$ for Ca, and $n=5$ for Sr.

\begin{table}
\caption[]{Intermediate dispersion coefficients for two interacting
alkaline-earth-metal atoms in the $(nsnp)\; \phantom{}^3P^o_2$ state.
See Eq.~(\ref{s5e3}) for a definition of $B_{J_1 J_2}$. The results
presented in Ref.~\cite{DePo03} have been converted using Eq.~(\ref{s5e4}).
All data are in atomic units.}
\label{tab8}
\begin{tabular}{cccc}
\hline
\hline
& Mg & Ca & Sr \\
\hline
$B_{1,1}$ (this work)          & $-35.6(2)$ & $-81(3)$ & $-139(7)$ \\
$B_{1,1}$ (Ref.~\cite{DePo03}) & $-38(4)$   & $-92(9)$ & $-158(16)$ \\
\hline
$B_{2,1}$ (this work)          & $42.5(2)$ & $119(5)$  & $196(9)$ \\
$B_{2,1}$ (Ref.~\cite{DePo03}) & $44(4)$   & $123(12)$ & $203(20)$ \\
\hline
$B_{2,2}$ (this work)          & $-51.9(2)$ & $-176(8)$  & $-280(10)$ \\
$B_{2,2}$ (Ref.~\cite{DePo03}) & $-52(5)$   & $-167(17)$ & $-264(26)$ \\
\hline
$B_{3,1}$ (this work)          & $-73.4(7)$ & $-203(10)$ & $-370(30)$ \\
$B_{3,1}$ (Ref.~\cite{DePo03}) & $-77(8)$   & $-225(23)$ & $-415(42)$ \\
\hline
$B_{3,2}$ (this work)          & $88.6(7)$ & $302(20)$ & $546(50)$ \\
$B_{3,2}$ (Ref.~\cite{DePo03}) & $90(9)$   & $306(31)$ & $555(56)$ \\
\hline
$B_{3,3}$ (this work)          & $-152(2)$  & $-553(70)$ & $-1210(200)$ \\
$B_{3,3}$ (Ref.~\cite{DePo03}) & $-156(16)$ & $-600(60)$ & $-1290(130)$ \\
\hline
\hline
\end{tabular}
\end{table}

The electric dipole--dipole dispersion physics is also more complicated
for $J=2$ than for $J=0$. We introduce {\em intermediate} dispersion 
coefficients to characterize the dispersion interaction between 
nonspherical atoms~\cite{SaGr03}:
\begin{eqnarray}
\label{s5e3}
B_{J_1 J_2} & = & (-1)^{1+J_1-J_2} \frac{4}{9} \frac{1}{(2 J_m + 1)^2} \\
& & \times \sum_{n_1,n_2}
\frac{\left|\left<n_1\right.\parallel D \parallel \left.m\right>\right|^2
\left|\left<n_2\right.\parallel D \parallel \left.m\right>\right|^2}
{E_{n_1}+E_{n_2}-2 E_m}. \nonumber
\end{eqnarray}
In our specific case of two interacting $^3P^o_2$ atoms, $J_m=2$, and 
$\left|n_1\right>$, $\left|n_2\right>$ denote even-parity
states with angular momenta $J_1$ and $J_2$, respectively. Both 
$J_1$ and $J_2$ can vary between $1$, $2$, and $3$, so there are six
nontrivial intermediate dispersion coefficients. They are displayed
in Table~\ref{tab8}. (The remaining three coefficients can be obtained 
by simple sign changes, according to Eq.~(\ref{s5e3}).) Derevianko
{\em et al.}~\cite{DePo03} use intermediate dispersion coefficients
defined as 
\begin{equation}
\label{s5e4}
C_6^{J_1 J_2} = -\frac{27}{8} (2 J + 1)^2 B_{J_1 J_2}.
\end{equation}
Their converted {\em ab initio} data are also shown in Table~\ref{tab8}.
The agreement between both data sets is satisfying.

The restriction to long-range expansion terms up to $1/R^6$ is valid only
for interatomic distances of the order of $100$ Bohr radii or larger. In this
regime, however, energies associated with rotations of the diatomic frame 
of an interacting pair are comparable with the interatomic interaction 
energies, and the coupling between the rotational and the electronic degrees
of freedom induced by the anisotropic interatomic interaction must
be explicitly taken into account. Reference~\cite{SaGr03} develops a
tensorial theory appropriate for this purpose and shows how the parameters 
${\cal Q}$ and $B_{J_1 J_2}$ enter the theory.

\section{Conclusion}
\label{sec6}

In this paper, we have theoretically investigated properties
of the energetically lowest $^3P^o$ manifold in the alkaline-earth-metal
atoms Mg, Ca, and Sr. The latter two, in particular, are the subject of strong
current interest. Using an electron-correlation approach for the valence shell
in combination with a physically well-motivated potential for the
valence interaction with the core shells, we have been able to reproduce
existing {\em ab initio} data with good accuracy. This provides 
an independent check on a number of crucial atomic parameters for present-day
experiments.

The calculations presented here also supply data that have not, to our 
knowledge, been presented elsewhere. First, the E1M1 decay rate of the 
$^3P^o_0$ state in spinless alkaline-earth-metal species was not previously
known. The difference in level structure to heliumlike atoms may motivate 
a direct search for the exotic E1M1 decay process in heavy magnesiumlike 
(or berylliumlike) ions. Second, our calculation of the hyperfine-induced 
E1 decay rate of the $^3P^o_0$ state in $^{43}$Ca, and also our determination 
of the magic wavelength for the corresponding clock transition from the 
ground state of calcium, may be of interest for the development of a future 
generation of optical atomic clocks. Finally, we have calculated the $C_6$ 
dispersion coefficient describing the long-range interaction between two 
metastable $^3P^o_0$ alkaline-earth-metal atoms. This system, which is free
from any losses due to Penning ionization, may be ideal for studying cold
collisions between spherical, metastable atoms.

\acknowledgments
R.S. gratefully acknowledges partial support by the Emmy Noether 
program of the German Research Foundation (DFG). This work was supported 
in part by the U.S. Department of Energy, Office of Science.

\end{document}